# Atomistic modeling of extreme near-field heat transport across nanogaps between two polar dielectric materials

Yangyu Guo [1, 3*], Mauricio Gómez Viloria [2], Riccardo Messina [2], Philippe Ben-Abdallah [2], and Samy Merabia [3†]

[1] *School of Energy Science and Engineering, Harbin Institute of Technology, Harbin 150001, China*

[2] *Laboratoire Charles Fabry, UMR 8501, Institut d'Optique, CNRS, Université Paris-Saclay, 2 Avenue Augustin Fresnel, 91127 Palaiseau Cedex, France*

[3] *Institut Lumière Matière, Université Claude Bernard Lyon 1-CNRS, Université de Lyon, Villeurbanne 69622, France*

(Dated: May 23, 2023)

## Abstract

The understanding of extreme near-field heat transport across vacuum nanogaps between polar dielectric materials remains an open question. In this work, we present a molecular dynamic simulation of heat transport across MgO-MgO nanogaps, together with a consistent comparison with the continuum fluctuational-electrodynamics theory using local dielectric properties. The dielectric function is computed by Green-Kubo molecular dynamics with the anharmonic damping properly included. As a result, the direct atomistic modeling shows significant deviation from the continuum theory even up to a gap size of few nanometers due to non-local dielectric response from acoustic and optical branches as well as phonon tunneling. The lattice anharmonicity is demonstrated to have a large impact on the energy transmission and thermal conductance, in contrast to its moderate effect reported for metallic vacuum nanogaps. The present work thus provides further insight into the physics of heat transport in the extreme near-field regime between polar materials, and put forward a methodology to account for anharmonic effects.

---

* yyguo@hit.edu.cn
† samy.merabia@univ-lyon1.fr

# 1. Introduction

When the gap between two objects is smaller than the thermal wavelength of photons (~10 $\mu$m at 300 K), the radiative heat transfer enters the near-field regime, where the heat flux can be several orders of magnitude higher than that predicted by the Stefan-Boltzmann law [1,2]. The near-field radiative heat transfer (NFRHT) has been well described by the classical fluctuational-electrodynamics (FE) theory [3], as widely verified by numerous experiments during the past decades [4-14]. However, it remains still an open question to fully understand the physics which drives the heat exchanges in extreme near-field regime, when the separation gap becomes smaller than few nanometers, corresponding to the transition regime between radiation and conduction. There have been very few experimental studies in such a regime [15-19]. Those studies were limited to the study of heat transfer between metals and they yield conflicting conclusions which are still in debate.

The investigation of extreme near-field heat transport is of vital significance in many applications such as near-field scanning thermal microscopy [20-22], heat-assisted magnetic recording [23-25], nano-photolithography [26] or non-contact friction [27,28]. On the other hand, it is of fundamental interest due to the emerging novel physics at such small scale including: *i*) the tunneling of phonons [29-32] and electrons [33,34] across a vacuum nanogap, and *ii*) the non-local dielectric response of materials [35,36]. Atomistic modeling becomes an essential tool to uncover the underlying physics, and also provide a reference to examine the validity of the continuum FE theory. Very recently, phonon tunneling across vacuum nanogaps between two infinite metallic plates has been studied by atomistic methods [37-40]. Similar atomistic studies have been conducted on heat transport across Si nanogaps using either empirical potential [41] or first-principles calculations [42]. All these works show that the phonon thermal conductance decays very rapidly as the gap size increases due to the relatively short range of atomic interactions in metallic and apolar solid systems. Generally the contribution of phonon tunneling becomes negligible beyond a gap size of 1 nm, beyond which the photonic contribution dominates [40,42]. Yet, very few studies have considered heat transfer across nanogaps between polar materials from an atomistic point of view.

In this work, we focus our attention on the extreme near-field heat transport between polar materials across vacuum gaps much thinner than the range of atomic interactions. A molecular-dynamics (MD) simulation of heat transport between two silica nano-particles was shown [33] to agree with the dipole-dipole interaction model after a separation distance of few diameters, where the thermal conductance ($h$) decays with the gap size ($d$) as $h \propto d^{-6}$. Such power-law was later obtained by a harmonic non-equilibrium Green's function (NEGF) simulation of the same configuration [43], where a slower decay of thermal conductance $h \propto d^{-4}$ was further shown at smaller gap size between 4 Å and few diameters due to the surface charge-charge interaction. The electron-cloud overlap and transition to heat conduction was finally inferred below a gap size of 4 Å [43]. Recently the transition from NFRHT to heat conduction was also studied across a vacuum gap between two infinite NaCl plates [44] by coupling the Maxwell equations to harmonic NEGF simulation. The gap conductance predicted by NEGF has been demonstrated to recover that by the FE theory beyond a gap size of ~1 nm, below which appreciable deviation was observed [44]. However, the long-range Coulomb interaction forces as input into NEGF were obtained from the solution of Maxwell equations for a system of harmonically-oscillating charged ions. Thus the role of lattice anharmonicity is elusive in their comparison of harmonic NEGF with FE theory, the latter requiring the dielectric function including the anharmonic damping.

Therefore, the current work aims to present a MD simulation of extreme near-field heat transport across a vacuum nanogap between two infinite MgO plates. A more realistic description of the anharmonic dynamics of ions and their long-range Coulomb interaction is provided in MD using an empirical potential [45], which describes reasonably well the dielectric function of the material. Furthermore, we propose a more consistent comparison between the atomistic modeling and FE theory by supplementing the latter with local dielectric properties calculated from Green-Kubo MD. As a result, we demonstrate non-negligible deviation of atomistic modeling from FE theory up to a gap size of 2 nm, while a gradual recovery of FE theory is expected at larger size. Finally, the role of anharmonicity, which is naturally included in MD, is shown to have appreciable effect on the energy transmission across the nanogap. The remaining of this manuscript is organized as follows: the methodology of atomistic modeling will be introduced in Section 2, followed by a discussion of the results in Section 3, and the concluding remarks will be made in Section 4.

## 2. Methodology

In this section, we first introduce in Section 2.1 the MD model of the MgO-MgO vacuum nanogap. The method to extract the spectral thermal conductance for the comparison with FE theory is also explained. In Section 2.2, we present a summary of FE theory for NFRHT between two infinite plates, together with the Green-Kubo MD method to calculate the local dielectric properties as input.

### 2.1 Molecular-dynamics simulations and analysis

Heat transport across a parallel vacuum MgO-MgO nanogap is simulated by non-equilibrium MD (NEMD) as implemented in the open-source package LAMMPS [46], as shown in Figure 1. The nanogap is obtained by shifting half part of a bulk MgO crystal along the [1 0 0] direction. In NEMD, the nanogap is sandwiched between a hot thermostat and a cold thermostat, with two fixed-layer regions at both ends. Periodic boundary conditions are imposed along the three directions of the system. The following pairwise atomic interaction potential is employed [45]:

$$\phi_{ij} = \frac{q_i q_j}{r_{ij}} - \frac{C_i C_j}{r_{ij}^6} + f\left(B_i + B_j\right) \exp\left(\frac{A_i + A_j - r_{ij}}{B_i + B_j}\right), \quad (1)$$

where the distance between atom $i$ and atom $j$ is $r_{ij} = |\mathbf{r}_i - \mathbf{r}_j|$, $q_i$ is the effective charge of atom $i$, and $f$ = 4.184 kJ/Å/mol, with other atomic parameters $A_i$, $B_i$ and $C_i$ given in Ref. [45]. Such a potential includes both long-range Coulomb interaction (the first term) and short-range bonding interaction (the second and third terms). This potential is adopted as it reproduces well the dielectric response of bulk MgO in the infrared regime [47], which is crucial for accurately describing the near-field heat transport across a vacuum gap. The particle-particle particle-mesh (pppm) method is implemented for the treatment of long-range Coulomb force with a cut-off radius of 10 Å for the direct interactions in real space.

Four gap sizes from 6 Å to 2 nm are considered, and the detailed dimensions of the molecular dynamic model after careful size independence verification are provided in Table 1. The length of the fixed-layer region ($L_f$) should be sufficiently large to avoid direct long-range interaction between the thermostats through the periodic boundary condition in the transport direction. In addition, the length of the device region on one side of the nanogap ($L_d$) should be sufficiently large to avoid direct long-range interaction between the thermostat on this side and the device region on the other side. A more detailed description of the size independence verification is given in Appendix. During the NEMD simulation, a time step of 0.5 fs is adopted. Firstly 1 million time steps are run to relax the whole system under the *NPT* (isothermal-isobaric) ensemble. Then, the fixed-layer regions are fixed and 2 million time steps are run to make the remaining free part reach a steady state under the effect of Langevin thermostats in the *NVE* (microcanonical) ensemble. Finally 4 million time steps of steady-state runs are performed for the calculation of the spectral and the overall thermal conductance of the nanogap. Five independent NEMD simulations are conducted for each gap size to reduce the statistical fluctuations.

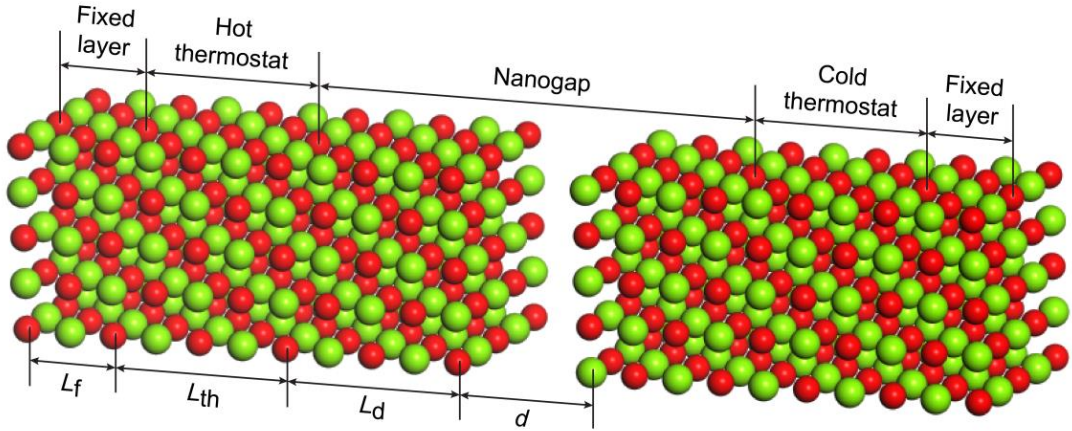

Figure 1. Schematic of MD model of MgO-MgO nanogap with a gap size *d*. The red and green atoms denote the $Mg^{2+}$ and $O^{2-}$ ions respectively. The lengths of the fixed layer and the thermostat are $L_f$ and $L_{th}$ respectively, whereas the length of device region on one side of the nanogap is $L_d$. The system is symmetric in terms of size. Periodic boundary conditions are applied along the three directions of the system.

Table 1. Dimension (in uc, i.e. conventional unit cell of MgO) of the MD model of MgO-MgO nanogap shown in Figure 1.

| Gap size $d$ (nm) | $L_f$ (uc) | $L_{th}$ (uc) | $L_d$ (uc) | Cross-section (uc × uc) |
|---|---|---|---|---|
| 0.6 | 8 | 5 | 8 | 8 × 8 |
| 1.0 | 12 | 5 | 12 | 8 × 8 |
| 1.2 | 12 | 5 | 12 | 8 × 8 |
| 2.0 | 14 | 5 | 14 | 10 × 10 |

The total heat flow from one side ($I$) of the nanogap device region to the other side ($J$) can be decomposed into its spectral component as:

$$Q_{I \to J} = \int_0^\infty q(\omega) \frac{d\omega}{2\pi}, \tag{2}$$

with the spectral heat flow computed from the Fourier transform of the time correlation function between atomic force and velocity [48,49]:

$$q(\omega) = 2\operatorname{Re} \sum_{\substack{i \in I \\ j \in J}} \int_{-\infty}^{\infty} \langle \mathbf{F}_{ji}(t) \cdot \mathbf{v}_j(0) \rangle \exp(i\omega t) dt, \tag{3}$$

where 'Re' denotes the real part, $\mathbf{F}_{ji}$ denotes the force on atom $j$ due to atom $i$, with $\mathbf{v}_j$ the atomic velocity, and the bracket '$\langle\ \rangle$' represents the non-equilibrium ensemble average as calculated by time average. In our implementation, we slightly transform Eq. (3) into the following form to avoid the huge storage of all the interatomic forces:

$$q(\omega) = 2\operatorname{Re} \sum_{j \in J} \int_{-\infty}^{\infty} \langle \mathbf{F}_j(t) \cdot \mathbf{v}_j(0) \rangle \exp(i\omega t) dt, \tag{4}$$

with $\mathbf{F}_j(t) = \sum_{i \in I} \mathbf{F}_{ji}(t)$ the overall force on atom $j$ in region $J$ from all the atoms in region $I$.

With the spectral heat flow obtained from NEMD simulation via Eq. (4), the transmission function across the nanogap is computed as [48]:

$$\Xi(\omega) = \frac{q(\omega)}{k_B \Delta T}, \tag{5}$$

where $k_B$ is the Boltzmann constant, and $\Delta T$ is the temperature difference between the hot and cold thermostats. Based on Landauer's formula, the classical and quantum thermal conductances of the nanogap can be computed respectively by [40]:

$$h = \frac{1}{A_c} \int_0^\infty k_B \Xi(\omega) \frac{d\omega}{2\pi}, \quad (6)$$

$$h = \frac{1}{A_c} \int_0^\infty \hbar\omega \frac{\partial f_{BE}(\omega)}{\partial T} \Xi(\omega) \frac{d\omega}{2\pi}. \quad (7)$$

In Eq. (6) and Eq. (7), the classical ($k_B$) and quantum heat capacity [$\hbar\omega \partial f_{BE}(\omega)/\partial T$] are used respectively, with $\hbar$ the reduced Planck constant, and $f_{BE}(\omega)$ the Bose-Einstein equilibrium distribution, and $A_c$ the cross-section area of the nanogap.

*2.2 Fluctuational-electrodynamic theory with microscopic inputs*

In parallel, we will compare the NEMD result with the prediction of FE theory using MD inputs. The NFRHT has been well described by the FE theory, which predicts the net heat flux between two infinite parallel plates at $T_1$ and $T_2$ respectively as [2,50]:

$$\Phi = \frac{1}{4\pi^2} \int_0^\infty \int_0^\infty [\Theta(\omega,T_1) - \Theta(\omega,T_2)] \xi_{12}(\omega,k_\parallel) k_\parallel dk_\parallel d\omega, \quad (8)$$

where $\Theta(\omega,T) = \hbar\omega[f_{BE}(\omega,T) + 1/2]$ is the mean energy of Planck oscillator at equilibrium, $k_\parallel$ is the parallel component of the photon wave vector and $\xi_{12}(\omega,k_\parallel)$ is the photon tunneling probability. In the limit of small temperature difference, Eq. (8) can be rewritten into the form of Landauer's formula [50], where the thermal conductance is calculated as:

$$h = \int_0^\infty \hbar\omega \frac{\partial f_{BE}}{\partial T} \left[ \int_0^\infty \xi_{12}(\omega,k_\parallel) k_\parallel \frac{dk_\parallel}{2\pi} \right] \frac{d\omega}{2\pi}. \quad (9)$$

The integrand in Eq. (9) thus represents the spectral thermal conductance of NFRHT, and will be compared to that obtained from NEMD in the previous sub-section.

The calculation of the photon tunneling probability $\xi_{12}(\omega,k_\parallel)$ requires the knowledge of the local dielectric function of bulk MgO [2,50]. To ensure a consistent comparison between the present NEMD and the FE theory, we compute the dielectric function by equilibrium MD (EMD) simulation using the same atomic interaction potential as in NEMD.

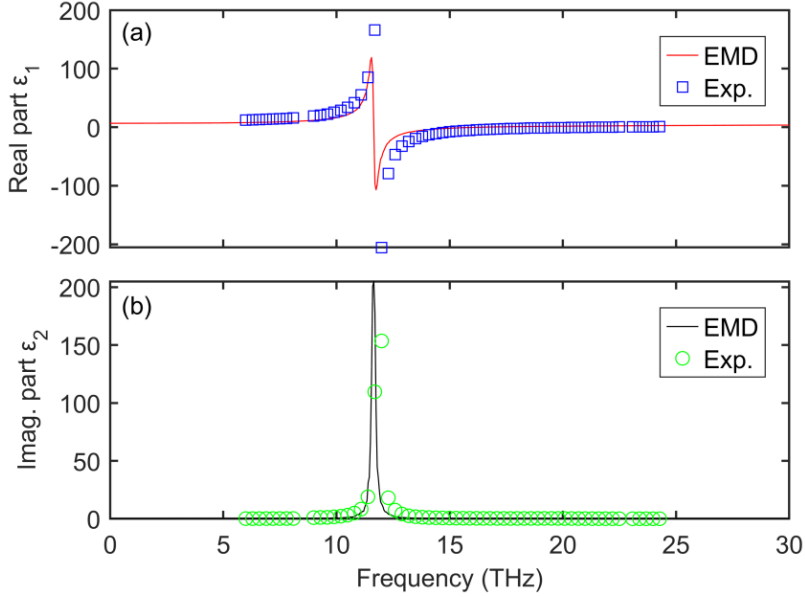

Figure 2. Local dielectric function ($\varepsilon = \varepsilon_1 + i\varepsilon_2$) of bulk MgO at 300 K in the infrared regime: (a) real part ($\varepsilon_1$), (b) imaginary part ($\varepsilon_2$). The discrete symbols represent the experimental data from the literature [51], whereas the solid lines denote the present result calculated by equilibrium molecular dynamics (EMD).

The local dielectric function tensor is related to the dielectric susceptibility $\chi_{\alpha\beta}(\omega)$ as [52,53]: $\varepsilon_{\alpha\beta}(\omega) = \delta_{\alpha\beta} + \chi_{\alpha\beta}(\omega)$, with $\delta_{\alpha\beta}$ the Kronecker delta, and $\chi_{\alpha\beta}(\omega)$ calculated via the fluctuation-dissipation theorem (i.e. Green-Kubo formula) [52-54]:

$$\chi_{\alpha\beta}(\omega) = \frac{V}{\varepsilon_0 k_B T}\left[\langle P_\alpha(0) P_\beta(0)\rangle + i\omega \int_0^\infty e^{i\omega t} \langle P_\alpha(t) P_\beta(0)\rangle dt\right], \quad (10)$$

where $V$ is the system volume, $\varepsilon_0$ is the vacuum permittivity. In Eq. (10), the polarization of the system is calculated as the density of dipole moment: $\mathbf{P}(t) = 1/V \sum_i q_i \mathbf{u}_i(t)$, with $\mathbf{u}_i = \mathbf{r}_i - \mathbf{r}_{i,0}$ the atomic displacement with respect to its equilibrium position $\mathbf{r}_{i,0}$. In the numerical implementation, we adopt a $10 \times 10 \times 10$ supercell of 8000 atoms in the EMD simulation with a time step of 0.5 fs. The size of the supercell has been tested to be sufficiently large to capture well the long-range interaction. Firstly, one million time steps are run under the *NVT* (canonical) ensemble for structure relaxation, after which 5 million time steps are run under the *NVE* ensemble. The polarization of the system is output once per 20 time steps during the *NVE* run. Ten independent simulations are conducted to reduce the statistical fluctuations. As shown in Figure 2, the local dielectric function calculated by EMD at 300 K generally agrees well with the available experimental data [51], similar to the findings in Ref. [47]. Note that the electronic degrees of freedom are not taken into account in MD simulation, which only captures the ionic contribution to dielectric response as the dominant mechanism in the infrared regime. Following Ref. [53], in Figure 2, we have included a constant correction term to account for the electronic contribution as: $\varepsilon_{\alpha\beta}(\omega) = \delta_{\alpha\beta} + \chi_{\alpha\beta}(\omega) + (\varepsilon_\infty - 1)\delta_{\alpha\beta}$, with $\varepsilon_\infty = 3.01$ being the high-frequency dielectric constant [51].

## 3. Results and Discussions

In this section, we first present the results of the spectral and overall thermal conductance of MgO-MgO nanogap at 300 K in Section 3.1 and Section 3.2, respectively. We also compare the results of atomistic modeling with FE theory. The role of lattice anharmonicity on heat transfer across the vacuum nanogap is finally discussed in Section 3.3.

### 3.1 Transmission and spectral thermal conductance

The frequency-dependent transmission functions across three MgO-MgO nanogaps with a gap size of 6 Å, 1 nm and 2 nm by NEMD at 300 K are shown in Figure 3(a). To have an intuitive understanding of which phonon branch contributes to the heat tunneling, we calculate the phonon dispersion via the non-local dielectric spectrum by EMD, as plotted in Figure 3(b). The Γ-X direction is chosen corresponding to the [1 0 0] transport direction. The frequency- and wave-vector-dependent (i.e. non-local) dielectric function is related to the dielectric susceptibility as: $\varepsilon_{\alpha\beta}(\omega, \mathbf{k}) = \varepsilon_\infty \delta_{\alpha\beta} + \chi_{\alpha\beta}(\omega, \mathbf{k})$, where $\chi_{\alpha\beta}(\omega, \mathbf{k})$ is calculated through a generalized version of the Green-Kubo formula in Eq. (10):

$$\chi_{\alpha\beta}(\omega, \mathbf{k}) = \frac{V}{\varepsilon_0 k_B T} \left[ \langle P_\alpha^*(0, \mathbf{k}) P_\beta(0, \mathbf{k}) \rangle + i\omega \int_0^\infty e^{i\omega t} \langle P_\alpha^*(t, \mathbf{k}) P_\beta(0, \mathbf{k}) \rangle dt \right], \tag{11}$$

with the superscript '*' denoting the complex conjugate. The wave-vector-dependent polarization of the system is calculated by the following projection:

$$\mathbf{P}(t, \mathbf{k}) = \frac{1}{V} \sum_{l\kappa} q_{l\kappa} \mathbf{u}_{l\kappa}(t) \exp(i\mathbf{k} \cdot \mathbf{r}_{l,0}), \tag{12}$$

where the atomic index '$l,\kappa$' includes the indices of lattice unit cell ($l$) and atoms within one unit cell ($\kappa$) respectively. The different phonon branches are clearly seen in Figure 3(b), with the broadening from the anharmonic phonon-phonon scattering. The present way to compute the phonon dispersion from EMD is in principle equivalent to the classical spectral energy density (SED) analysis used to extract the phonon dispersion and lifetimes [55].

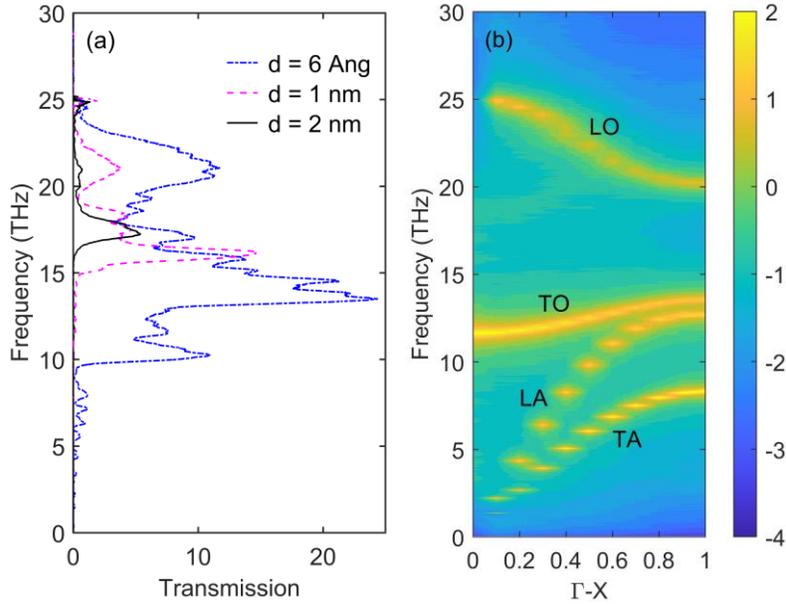

Figure 3. Frequency-dependent (a) transmission function by NEMD of MgO-MgO nanogaps with gap size of 6 Å, 1 nm and 2 nm respectively, and (b) imaginary part of the non-local dielectric function in log scale (i.e., $\log \varepsilon_2$) of bulk MgO showing the phonon dispersion along the Γ-X direction. The system temperature is 300 K.

As shown in Figure 3(a), the transmission function is generally reduced in both spectral range and magnitude as the gap size increases. For gap size > 1 nm, the transmission function is almost negligible at a frequency lower than 15 THz. In other words, only optical modes could go through those nanogaps, based on the phonon dispersion in Figure 3(b). At a gap size of 6 Å, there is also energy transmission in the lower frequency range corresponding to acoustic phonons. To be more specific, the significant peak around 10 THz shall be contributed by the longitudinal acoustic (LA) phonons, whereas the few small peaks between 5~10 THz could be due to both LA and transverse acoustic (TA) phonons.

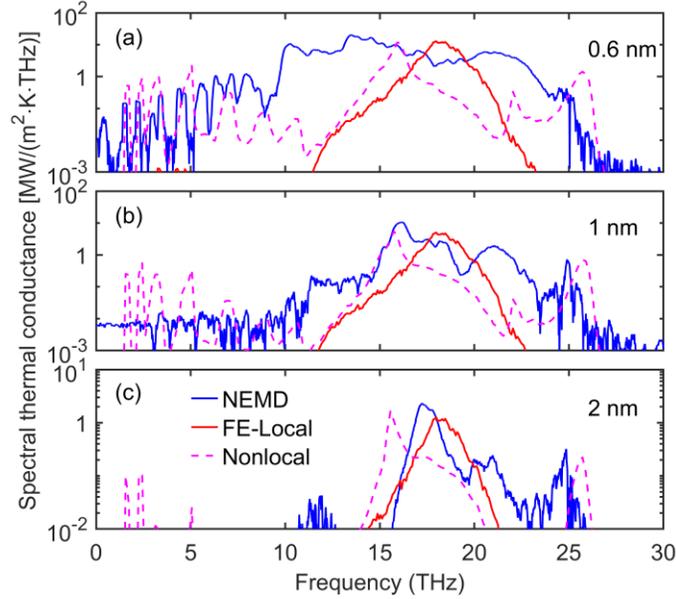

Figure 4. Spectral thermal conductance of MgO-MgO nanogaps at gap size of (a) 0.6 nm, (b) 1 nm and (c) 2 nm. The blue solid lines denote the present quantum-corrected result by non-equilibrium molecular dynamics (NEMD), whereas the red solid lines denote the result by fluctuational-electrodynamic (FE) theory with local dielectric function calculated by EMD (c.f. Figure 2), and the magenta dashed lines denote the result by non-local theory for radiative heat transfer from Ref. [36].

To have deeper insight, we make a comparison of the spectral thermal conductance by the atomistic modeling to that predicted by FE theory with local dielectric function, as shown in Figure 4. For a fair comparison, the quantum-corrected thermal conductance by NEMD as calculated in Eq. (7) is used, whereas the electronic contribution to the dielectric function ($\varepsilon_\infty$) is excluded in the FE theoretical calculation. There is only one peak in the thermal spectrum of NRFHT predicted by FE-local theory. This resonance peak is known to be caused by the surface phonon polaritons from the hybridization of transverse optical (TO) phonons and electromagnetic waves [56]. The resonance frequency lies where the real part of the local dielectric function in Figure 2 reaches -1. At the smallest gap size of 6 Å in Figure 4(a), a very broad thermal spectrum is obtained by NEMD, which shows a very large deviation from the FE-local result. This is a clear evidence of the failure of the continuum FE theory with local dielectric response. As the gap size increases to 1 nm and 2 nm in Figure 4(b) and (c) respectively, the difference between the results of NEMD and FE-local theory tends to become gradually smaller. Actually, at the 2 nm gap, the thermal spectrum from NEMD shows a significant peak at ~17.5 THz that is quite close to the resonance peak predicted by FE theory. Note that the height of the second peak at ~25 THz is only ~10% of that of the main peak. We have not simulated even larger nanogap due to the intensive computational cost required to accurately describe the slowly-decaying long-range Coulomb forces.

The deviation between continuum theory and atomistic modeling at very small gap size has two main origins: *i*) the phonon tunneling due to direct short-range atomic interaction across the gap, similar to that in heat transport through metallic vacuum nanogaps [40]; *ii*) the non-local dielectric response of polar materials from long-range interaction between dipole moments generated by ionic vibrations, including the optical response of both acoustic phonons [36] and longitudinal optical (LO) phonons [57]. The non-local dielectric response is highlighted in the frequency- and wave-vector-dependent dielectric function shown in Figure 3(b). In contrast to local dielectrics where only Γ-point TO phonons couple to the electromagnetic waves, all the branches including LO, TO, LA and TA phonons throughout the Brillouin zone show clear optical response (i.e. infrared adsorption). As a comparison, we include in Figure 4 our recent result by non-local theory for radiative heat transfer at atomic scale [36]. Although a perfect agreement with non-local theory is not achieved, our NEMD simulation captures well some crucial features of non-local dielectric response predicted by the theory, especially the contribution from acoustic phonons at the smallest gap. It remains, however, a challenging task to disentangle the contributions from phonon tunneling and from non-local dielectric response in the present methodology.

*3.2 Thermal conductance*

The overall thermal conductance of MgO-MgO nanogap is obtained by integrating the spectral thermal conductance in Section 3.1 over the whole frequency range, as given in Figure 5(a) in log-log scale and in Figure 5(b) in log-normal scale. In the current studied range of gap size from 6 Å to 2 nm, the gap conductance by NEMD is higher than that of NFRHT predicted by FE theory with local dielectric function. The underlying reason is due to both phonon tunneling and non-local dielectric response, as discussed at the end of Section 3.1. On the other hand, the NEMD result is gradually approaching that of FE theory as the gap size increases, with a faster decaying trend of $h \propto d^{-2.8}$ compared to $h \propto d^{-2}$ in the latter. There is still non-negligible difference between them at the 2 nm gap, while the convergence of atomistic modeling to FE theory is expected at larger gap size. The critical gap size where the continuum theory is recovered is larger than that (~1 nm) found in a previous study of NaCl nanogaps [44]. A recent study [57] has estimated a nonlocal length of ~10 nm in polar dielectric materials, which indeed indicates appreciable non-local optical response at few-nanometers scale. This is caused by the slowly-decaying Coulomb interaction, which also explains the much slower decay trend of the thermal conductance compared to that ($h \propto d^{-9}$) characterizing the metallic nanogaps [40]. Note that in NEMD, the thermal conductance will not diverge as $h \propto d^{-2.8}$ when the gap size further decreases. Instead, a transition to the contact heat conduction should occur, as already shown in previous works [43,44]. Actually, when setting the gap size as half of the lattice constant of bulk MgO in NEMD simulation, we will recover exactly the heat conduction across a MgO film.

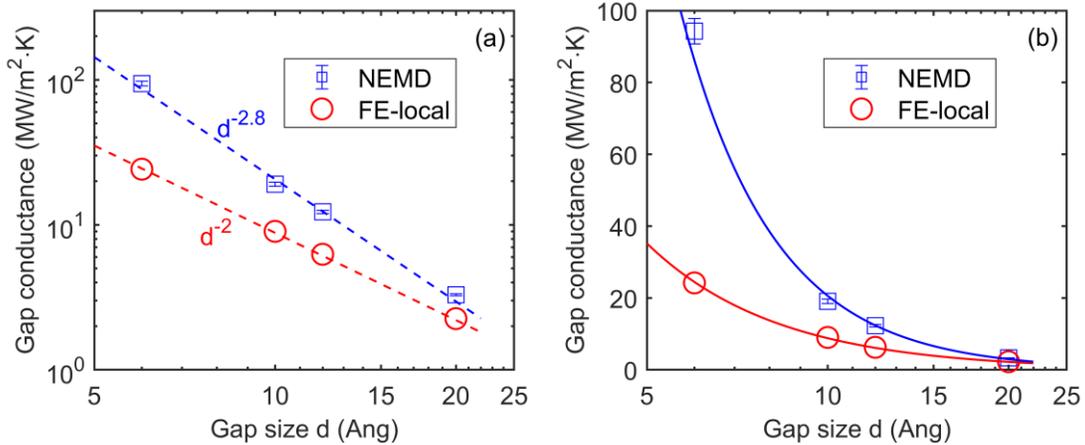

Figure 5. Thermal conductance of MgO-MgO nanogaps versus gap size $d$ at 300 K in (a) log-log scale, (b) log-normal scale. The blue squares with error bar denote the present quantum-corrected results by non-equilibrium molecular dynamics (NEMD), whereas the red circles denote the near-field radiative heat transfer result by FE theory using the local dielectric function calculated by EMD (c.f. Figure 2). The solid lines in (b) are used for guiding the eye.

We would like to point out the importance of taking into account the anharmonicity in the comparison of atomistic modeling and continuum FE theory. In lattice dynamic theory, the local dielectric function of simple ionic crystals like MgO is given by [58]:

$$\varepsilon(\omega) = \varepsilon_\infty + \frac{S}{\Omega_{TO}^2(\omega) - \omega^2 - 2i\omega_{TO}\Gamma(\omega)}, \quad (13)$$

where $\Omega_{TO}(\omega) = \omega_{TO} + \Delta(\omega)$ is the renormalized eigenfrequency of TO phonon considering its frequency shift $\Delta(\omega)$, $\Gamma(\omega)$ is its damping resulting from anharmonic phonon-phonon scattering, and $S$ is the oscillator strength. The imaginary part of the dielectric function in Eq. (13) is expressed as a Lorentzian function:

$$\varepsilon_2(\omega) = \frac{2S\omega_{TO}\Gamma(\omega)}{\left[\Omega_{TO}^2(\omega) - \omega^2\right]^2 + 4\omega_{TO}^2\Gamma^2(\omega)}. \quad (14)$$

In the harmonic limit [$\Gamma(\omega) \to 0$], the imaginary part in Eq. (14) will be reduced to a Dirac delta function, which makes the calculation of NFRHT via Eq. (9) impractical. From this perspective, the comparison of harmonic NEGF to FE theory for heat transport across polar dielectric nanogaps [44] remains debatable due to the inconsistency between the inputs of the atomic interaction forces. To be more specific, only harmonic forces are employed in the former whereas both the harmonic and anharmonic forces are essential for calculating the dielectric function in the latter. In this work, our comparison is more consistent since the same atomic interaction is included in the atomistic modeling and in the calculation of dielectric function as input into the continuum FE theory.

*3.3 The role of anharmonicity*

Finally we investigate the role of anharmonicity on the extreme near-field heat transport across MgO-MgO nanogap, as motivated by the significant effect of the anharmonic phonon-phonon scattering on the dielectric response of polar materials. The 1 nm nanogap is considered, and we raise the strength of anharmonicity by varying the system temperature from 50 K to 300 K. As shown in Figure 6(a), the classical thermal conductance obtained by NEMD increases as the temperature rises, and its value at 300 K is more than twice that at 50 K. It has been known that the increase of thermal boundary conductance with temperature at solid-solid interface predicted by NEMD is attributed to the inelastic anharmonic phonon scattering [59]. Thus the present temperature-dependent trend of gap conductance should be also caused by the lattice anharmonicity. As a comparison, the thermal conductance of metallic vacuum nanogaps in our previous study increases only ~20%-30% from 1 K to 300 K, which indicates moderate anharmonic effect [40]. According to Eq. (6), the increasing gap conductance could only come from the enhanced energy transmission across the gap, since the spectral heat capacity is a constant ($k_B$) in the classical limit. In Figure 6(b), the quantum-corrected thermal conductance calculated via Eq. (7) varies by more than four orders of magnitude in the same temperature range, as mainly contributed by the increasing phonon population from Bose-Einstein statistics.

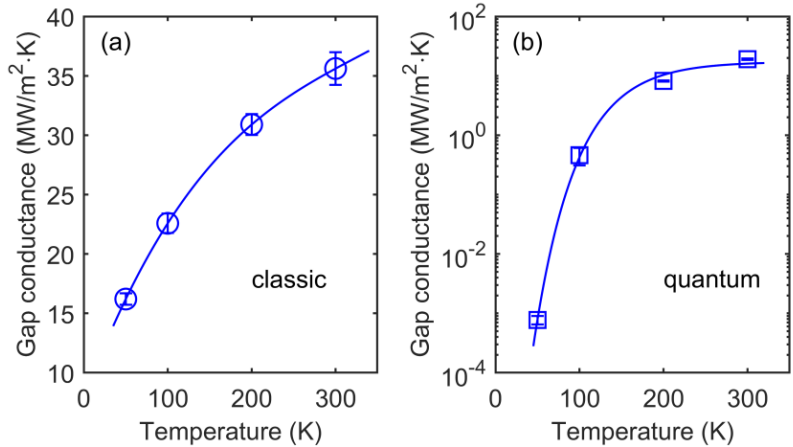

Figure 6. Temperature-dependent thermal conductance of 1nm MgO-MgO nanogap by NEMD: (a) classical conductance via Eq. (6), and (b) quantum-corrected conductance via Eq. (7). The solid lines are used for guiding the eyes.

The enhanced energy transmission across the nanogap with increasing temperature is explicitly demonstrated in Figure 7(a). Interestingly, the enhancement occurs mainly in the frequency range between 15 THz and 20 THz, while the transmission function is almost independent of temperature between 20 THz and 25 THz. As a comparison, we also plot the temperature-dependent transmission function of NFRHT predicted by the FE theory, i.e. the term in the square bracket of Eq. (9), as shown in Figure 7(b). Here the local dielectric functions at corresponding temperatures are computed by EMD and then used for FE theoretical calculation. As the temperature increases, the peak in the transmission function curve is reduced while it is broadened. This can be explained by the same trend of local dielectric function in Eq. (14) since the damping (i.e. the phonon linewidth) increases with temperature. However, the trend is quite different in the result by NEMD, namely, the peaks are enhanced and also broadened between 15 THz and 20 THz. It could be not explained by the effect of anharmonicity on the local dielectric response of the material. The impact of anharmonic phonon scattering on either the non-local response or the phonon tunneling may be one possible reason, which remains to be further investigated in future works.

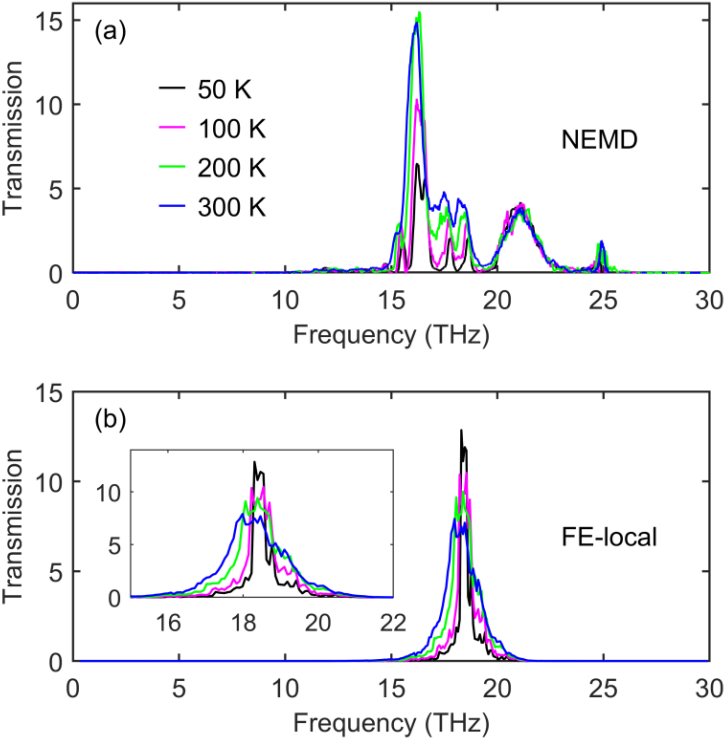

Figure 7. Temperature-dependent transmission function across MgO-MgO nanogap with a gap size of 1 nm: (a) NEMD result, (b) near-field radiative heat transfer result by fluctuational-electrodynamic (FE) theory with local dielectric function calculated by EMD (c.f. Figure 2). The inset in (b) shows the enlarged plot between 15 THz and 22 THz.

## 4. Conclusions

In summary, we present a multiscale framework to investigate the extreme near-field heat transport across polar dielectric nanogaps. At angstrom- and nanometer-sized gaps, the direct atomistic modeling shows significant deviation from the continuum fluctuational-electrodynamics theory with consistent microscopic input, due to both phonon tunneling and non-local dielectric response. The atomistic-modeling results gradually approach that of the continuum theory and are expected to recover the latter at larger nanogaps. The energy transmission across the nanogap increases as the temperature rises, attributed to the effect of lattice anharmonicity. Our results highlight the importance of non-local effects stemming from both acoustic and optical phonons in the extreme near-field regime. The atomistic approach allows to take into account all contributions to the energy transfer. However, we are not able to quantify the relative contribution of phonon tunneling with respect to non-local effect. This analysis will be the subject of a future work.

**Acknowledgements**

This work was supported by the French Agence Nationale de la Recherche (ANR), under Grant No. ANR-20-CE05-0021-01 (NearHeat). The simulation used the computational resources of Raptor of iLM at the Université Claude Bernard Lyon 1.

**Appendix. Size independence verification of molecular dynamic model**

Here we provide the details of size independence verification of the dimension of molecular dynamic model for the 1 nm MgO nanogap as an example. Firstly we verify the influence of cross-section size by fixing $L_{th}$ = 5 uc, $L_f = L_d$ = 4 uc. As the cross-section increases from 6 uc × 6 uc to 10 uc × 10 uc, the classical thermal conductance directly obtained by NEMD increases less than 3%, as summarized in Table 2. Thus we adopt a cross-section of 8 uc × 8 uc for the MD simulation of 1 nm gap.

Table 2. Independence verification of the cross-section size (in uc, i.e. conventional unit cell of MgO) in molecular dynamic model of 1 nm MgO-MgO nanogap. $L_{th}$ = 5 uc, $L_f = L_d$ = 4 uc. The standard deviation of the gap conductance is computed from five independent NEMD simulations.

| Cross-section (uc × uc) | $h_{NEMD}$ ($\times 10^7$ W/m$^2$·K) |
|---|---|
| 6 × 6 | 3.82 (± 0.06) |
| 8 × 8 | 3.83 (± 0.10) |
| 10 × 10 | 3.93 (± 0.04) |

We further verify the influence of the size of the fixed layer ($L_f$) and the device region ($2L_d$) at $L_{th}$ = 5 uc using a cross-section of 8 uc × 8 uc. As $L_f$ and $L_d$ increases, the classical thermal conductance directly obtained by NEMD ($h_{NEMD}$) almost does not change, as shown in Table 3. In contrast, the classical thermal conductance obtained by integrating the spectral heat current from NEMD ($h_{SHC}$), i.e. based on Eq. (6), gradually increases and approaches that directly obtained by NEMD. We have a small mismatch between $h_{SHC}$ and $h_{NEMD}$ since the spectral heat current is calculated between the left-hand side and right-hand side of the device region. Due to the long-range Coulomb interaction, there is some heat flow between the left-hand (right-hand) thermostat and right-hand (left-hand) side of the device region. To avoid such spurious effect and considering the periodic boundary condition along the transport direction, we increase $L_f$ and $L_d$ to be sufficiently large such that difference between $h_{SHC}$ and $h_{NEMD}$ is less than 5%, i.e. within the statistical uncertainty of MD simulation, as summarized in Table 3. Finally we adopt $L_f = L_d$ = 12 uc for the MD simulation of 1 nm gap.

Table 3. Independence verification of the size (in uc, i.e. conventional unit cell of MgO) of fixed layer and device region in molecular dynamic model of 1 nm MgO-MgO nanogap. $L_{th}$ = 5 uc, cross-section size: 8 uc × 8 uc. $h_{SHC}$ is calculated by integrating the spectral heat current (SHC) extracted from NEMD.

| $L_f$ (uc) | $L_d$ (uc) | $h_{NEMD}$ (×10$^7$ W/m$^2$·K) | $h_{SHC}$ (×10$^7$ W/m$^2$·K) | $h_{SHC} / h_{NEMD}$ (%) |
|---|---|---|---|---|
| 4 | 4 | 3.83 (± 0.10) | 3.37 (± 0.13) | 87.98 |
| 4 | 6 | 3.82 (± 0.05) | 3.50 (± 0.12) | 91.53 |
| 4 | 8 | 3.73 (± 0.08) | 3.47 (± 0.06) | 93.07 |
| 8 | 8 | 3.82 (± 0.08) | 3.57 (± 0.07) | 93.38 |
| 10 | 10 | 3.79 (± 0.13) | 3.58 (± 0.15) | 94.41 |
| 12 | 12 | 3.75 (± 0.11) | 3.56 (± 0.14) | 95.01 |

**References**

[1] A. Volokitin and B. N. Persson, Reviews of Modern Physics **79**, 1291 (2007).
[2] Z. M. Zhang, *Nano/microscale heat transfer* (McGraw-Hill New York, 2007).
[3] Sergei M. Rytov, Yurii A. Kravtsov, and V. I. Tatarskii, *Principles of Statistical Radiophysics 3* (Springer, Heidelberg, 1989).
[4] C. Hargreaves, Physics Letters A **30**, 491 (1969).
[5] A. Narayanaswamy, S. Shen, and G. Chen, Physical Review B **78**, 115303 (2008).
[6] S. Shen, A. Narayanaswamy, and G. Chen, Nano Letters **9**, 2909 (2009).

[7] E. Rousseau, A. Siria, G. Jourdan, S. Volz, F. Comin, J. Chevrier, and J.-J. Greffet, Nature Photonics **3**, 514 (2009).
[8] R. Ottens, V. Quetschke, S. Wise, A. Alemi, R. Lundock, G. Mueller, D. H. Reitze, D. B. Tanner, and B. F. Whiting, Physical Review Letters **107**, 014301 (2011).
[9] T. Kralik, P. Hanzelka, M. Zobac, V. Musilova, T. Fort, and M. Horak, Physical Review Letters **109**, 224302 (2012).
[10] B. Song, Y. Ganjeh, S. Sadat, D. Thompson, A. Fiorino, V. Fernández-Hurtado, J. Feist, F. J. Garcia-Vidal, J. C. Cuevas, and P. Reddy, Nature Nanotechnology **10**, 253 (2015).
[11] K. Kim, B. Song, V. Fernández-Hurtado, W. Lee, W. Jeong, L. Cui, D. Thompson, J. Feist, M. H. Reid, F. J. García-Vidal, J. C. Cuevas, E. Meyhofer, and P. Reddy, Nature **528**, 387 (2015).
[12] R. St-Gelais, L. Zhu, S. Fan, and M. Lipson, Nature Nanotechnology **11**, 515 (2016).
[13] M. Ghashami, H. Geng, T. Kim, N. Iacopino, S. K. Cho, and K. Park, Physical Review Letters **120**, 175901 (2018).
[14] H. Salihoglu, W. Nam, L. Traverso, M. Segovia, P. K. Venuthurumilli, W. Liu, Y. Wei, W. Li, and X. Xu, Nano Letters **20**, 6091 (2020).
[15] A. Kittel, W. Müller-Hirsch, J. Parisi, S.-A. Biehs, D. Reddig, and M. Holthaus, Physical Review Letters **95**, 224301 (2005).
[16] L. Worbes, D. Hellmann, and A. Kittel, Physical Review Letters **110**, 134302 (2013).
[17] K. Kloppstech, N. Könne, S.-A. Biehs, A. W. Rodriguez, L. Worbes, D. Hellmann, and A. Kittel, Nature Communications **8**, 14475 (2017).
[18] L. Cui, W. Jeong, V. Fernández-Hurtado, J. Feist, F. J. García-Vidal, J. C. Cuevas, E. Meyhofer, and P. Reddy, Nature Communications **8**, 14479 (2017).
[19] A. Jarzembski, T. Tokunaga, J. Crossley, J. Yun, C. Shaskey, R. A. Murdick, I. Park, M. Francoeur, and K. Park, Physical Review B **106**, 205418 (2022).
[20] Y. De Wilde, F. Formanek, R. Carminati, B. Gralak, P.-A. Lemoine, K. Joulain, J.-P. Mulet, Y. Chen, and J.-J. Greffet, Nature **444**, 740 (2006).
[21] I. Altfeder, A. A. Voevodin, and A. K. Roy, Physical Review Letters **105**, 166101 (2010).
[22] A. C. Jones and M. B. Raschke, Nano Letters **12**, 1475 (2012).
[23] M. H. Kryder, E. C. Gage, T. W. McDaniel, W. A. Challener, R. E. Rottmayer, G. Ju, Y.-T. Hsia, and M. F. Erden, Proceedings of the IEEE **96**, 1810 (2008).
[24] W. Challener, C. Peng, A. Itagi, D. Karns, W. Peng, Y. Peng, X. Yang, X. Zhu, N. Gokemeijer, and Y.-T. Hsia, Nature Photonics **3**, 220 (2009).
[25] B. C. Stipe, T. C. Strand, C. C. Poon, H. Balamane, T. D. Boone, J. A. Katine, J.-L. Li, V. Rawat, H. Nemoto, and A. Hirotsune, Nature Photonics **4**, 484 (2010).
[26] W. Srituravanich, N. Fang, C. Sun, Q. Luo, and X. J. N. l. Zhang, Nano Letters **4**, 1085 (2004).
[27] M. Lee, R. L. Vink, C. A. Volkert, and M. Krüger, Physical Review B **104**, 174309 (2021).
[28] A. Volokitin, Applied Surface Science Advances **6**, 100160 (2021).
[29] M. Prunnila and J. Meltaus, Physical Review Letters **105**, 125501 (2010).
[30] Y. Ezzahri and K. Joulain, Physical Review B **90**, 115433 (2014).
[31] J. Pendry, K. Sasihithlu, and R. Craster, Physical Review B **94**, 075414 (2016).
[32] Z. Geng and I. J. Maasilta, Physical Review Research **4**, 033073 (2022).
[33] Z.-Q. Zhang, J.-T. Lü, and J.-S. Wang, Physical Review B **97**, 195450 (2018).
[34] M. Gómez Viloria, Y. Guo, S. Merabia, P. Ben-Abdallah, and R. Messina, Physical Review B **107**, 125414 (2023).
[35] P.-O. Chapuis, S. Volz, C. Henkel, K. Joulain, and J.-J. Greffet, Physical Review B **77**, 035431 (2008).
[36] M. Gómez Viloria, Y. Guo, S. Merabia, R. Messina, and P. Ben-Abdallah, arXiv, 2302.00520 (2023).

[37] A. Alkurdi, C. Adessi, F. Tabatabaei, S. Li, K. Termentzidis, and S. Merabia, International Journal of Heat Mass Transfer **158**, 119963 (2020).
[38] T. Tokunaga, A. Jarzembski, T. Shiga, K. Park, and M. Francoeur, Physical Review B **104**, 125404 (2021).
[39] W. Chen and G. Nagayama, International Journal of Heat Mass Transfer **176**, 121431 (2021).
[40] Y. Guo, C. Adessi, M. Cobian, and S. Merabia, Physical Review B **106**, 085403 (2022).
[41] D. P. Sellan, E. Landry, K. Sasihithlu, A. Narayanaswamy, A. McGaughey, and C. Amon, Physical Review B **85**, 024118 (2012).
[42] T. Tokunaga, M. Arai, K. Kobayashi, W. Hayami, S. Suehara, T. Shiga, K. Park, and M. Francoeur, Physical Review B **105**, 045410 (2022).
[43] S. Xiong, K. Yang, Y. A. Kosevich, Y. Chalopin, R. D'Agosta, P. Cortona, and S. Volz, Physical Review Letters **112**, 114301 (2014).
[44] V. Chiloyan, J. Garg, K. Esfarjani, and G. Chen, Nature Communications **6**, 6755 (2015).
[45] M. Matsui, The Journal of Chemical Physics **91**, 489 (1989).
[46] A. P. Thompson, H. M. Aktulga, R. Berger, D. S. Bolintineanu, W. M. Brown, P. S. Crozier, P. J. in't Veld, A. Kohlmeyer, S. G. Moore, and T. D. Nguyen, Computer Physics Communications **271**, 108171 (2022).
[47] Y. Chalopin, M. Hayoun, S. Volz, and H. Dammak, Applied Physics Letters **104**, 011905 (2014).
[48] K. Sääskilahti, J. Oksanen, J. Tulkki, and S. Volz, Physical Review B **90**, 134312 (2014).
[49] K. Sääskilahti, J. Oksanen, S. Volz, and J. Tulkki, Physical Review B **91**, 115426 (2015).
[50] S.-A. Biehs, E. Rousseau, and J.-J. Greffet, Physical Review Letters **105**, 234301 (2010).
[51] E. D. Palik, *Handbook of Optical Constants of Solids II* (Academic Press, San Diego, 1998), Vol. II.
[52] A. Maradudin and R. J. P. R. Wallis, Physical Review **123**, 777 (1961).
[53] F. Gangemi, A. Carati, L. Galgani, R. Gangemi, and A. Maiocchi, Europhysics Letters **110**, 47003 (2015).
[54] W. Chen and L.-S. Li, Journal of Applied Physics **129**, 244104 (2021).
[55] J. A. Thomas, J. E. Turney, R. M. Iutzi, C. H. Amon, and A. J. McGaughey, Physical Review B **81**, 081411 (2010).
[56] J.-P. Mulet, K. Joulain, R. Carminati, and J.-J. Greffet, Microscale Thermophysical Engineering **6**, 209 (2002).
[57] C. R. Gubbin and S. De Liberato, Physical Review X **10**, 021027 (2020).
[58] P. Brüesch, *Phonons: Theory and Experiments II* (Springer-Verlag, Heidelberg, 1986).
[59] R. J. Stevens, L. V. Zhigilei, and P. M. Norris, International Journal of Heat Mass Transfer **50**, 3977 (2007).